\documentclass[preprint,aps,pra,showpacs,floatfix]{revtex4-1}
\usepackage{amsmath, amssymb}
\usepackage{times}
\usepackage {euscript}
\usepackage{graphicx}
\usepackage{amsthm}
\usepackage{epsfig}
\usepackage{bm}
\usepackage{multirow}
\usepackage{longtable}

\usepackage[dvipsnames]{xcolor}

\definecolor{newgreen}{HTML}{32CD32}

\newcommand{\bp}{{\bf p}}
\newcommand{\bk}{{\bf k}}
\newcommand{\br}{{\bf r}}
\newcommand{\bn}{{\bf n}}
\newcommand{\bb}{{\bf b}}
\newcommand{\balpha}{\boldsymbol{\alpha}}
\newcommand{\bepsilon}{\boldsymbol{\epsilon}}
\graphicspath{{graph/}}
\begin{document}
\thispagestyle{empty}
\title{
Radiative recombination of twisted electrons with bare nuclei:\\
going beyond the Born approximation
}
\author{V.~A.~Zaytsev$^{1, 2}$,
        V.~G.~Serbo$^{3, 4}$,
        and V.~M.~Shabaev$^{1}$}
\affiliation{
$^1$ Department of Physics, St. Petersburg State University,
     7/9 Universitetskaya naberezhnaya, St. Petersburg 199034, Russia \\
$^2$ ITMO University,
     Kronverkskii ave 49, 197101 Saint Petersburg, Russia         \\
$^3$ Novosibirsk State University,
     RUS--630090, Novosibirsk, Russia     \\
$^4$ Sobolev Institute of Mathematics,
     RUS--630090, Novosibirsk, Russia
\vspace{10mm}
}
%
\begin{abstract}
We present a fully relativistic investigation of the 
radiative recombination of a twisted electron with a bare heavy nucleus.
The twisted electron is described by the wave function which accounts 
for the interaction with the nucleus in all orders in $\alpha Z$.
We use this wave function to derive the probability of the radiative 
recombination with a single ion being shifted from the twisted electron
propagation direction.
We also consider more realistic experimental scenarios where the 
target is either localized (mesoscopic) or infinitely wide (macroscopic).
The situation when the incident electron is a coherent superposition of 
two vortex states is considered as well.
For the nonrelativistic case we present analytical
expressions which support our numerical calculations. 
We study in details the influence of the electron twistedness on the 
polarization and angular distribution of the emitted photon.
It is found that these properties of the outgoing photon might be very 
sensitive to the total angular momentum and kinematic properties of 
twisted beams. 
Therefore, the recombination of the twisted electrons can serve as a 
valuable tool for atomic investigations as well as for the diagnostics of 
the vortex electron beams.
\end{abstract}
%
\maketitle
%
%
%
%
%
%
\section{INTRODUCTION}
%
%
Since the theoretical prediction~\cite{Bliokh_PRL99_190404:2007}, the
twisted (or vortex) electrons have become one of the most attractive
objects of interest in the contemporary physics.
They are characterized by the energy $\varepsilon$, one of the 
momentum components $p_z$ which sets the propagation direction, 
and the projection of the total angular momentum $\hbar m$ on this 
direction.
The interest to such particles is caused mainly by the non-zero value 
of this projection, being an additional degree of freedom.
Moreover, the growth of $m$ leads to the increase of the twisted 
electron magnetic moment $\mu = m\mu_{B}$ ($\mu_{B}$ is the Bohr 
magneton) along the propagation direction.
This fact points on the sensitivity of the electron vortex beams to 
magnetic properties of matter~\cite{Rusz_PRL111_105504:2013, Beche_NP10_26:2014,
Schattschneider_U136_81:2014, Edstrom_PRL116_127203:2016}.
First experimental realizations of these electrons were performed just 
half a decade ago~\cite{Verbeeck_N467_301:2010, Uchida_N464_737:2010, 
McMorran_S331_192:2011}.
In these experiments the twisted electrons possessing $m = 50$ were 
obtained.
Presently, the twisted electrons with the momentum projection $m$ up to 
$500$ can be routinely produced at electron microscopes~%
\cite{Grillo_PRL114_034801:2015, Mafakheri_MM21_667:2015}.
Electrons with such a large value of the total angular momentum 
projection can be used for the detection of the polarization radiation~%
\cite{Ivanov_PRL110_264801:2013, Ivanov_PRA88_043840:2013}.
In addition, the vortex electrons provide a new opportunity to get a 
deeper insight in the role of the spin-orbit interaction in various 
atomic processes.
%
%
\\ \indent
%
%
Despite a great interest, there are only few works presented in the
literature being dedicated to the investigation of the processes
involving ionic (or atomic) targets and twisted electrons~%
\cite{Boxem_PRA89_032715:2014, Matula_NJP16_053024:2014,
Boxem_PRA91_032703:2015, Serbo_PRA92_012705:2015}.
In all these articles the interaction of the twisted electrons with
targets was considered perturbatively in the
framework of the first Born approximation.
This approximation stays valid only for light systems with relatively
small nuclear charge $Z$ and at rather large projectile velocities.
Meanwhile the manifestation of the twistedness is expected to become the most
pronounced in heavy systems where the spin-orbit interaction increases
drastically.
In order to investigate the processes involving heavy systems one needs
to account for the interaction of twisted electrons with the targets in
all orders in $\alpha Z$.
This can be achieved via the construction of the twisted electron
relativistic wave function in the long-range Coulomb field of the nucleus.
In the present paper, we construct such a wave function and utilize it
for the description of the radiative recombination (RR) of a twisted
electron with a bare heavy nucleus.
Two types of the targets are considered, namely the infinitely extended one
(macroscopic) and the target with a finite spatial distribution 
(mesoscopic).
For the macroscopic target, we compare our nonrelativistic 
results with the ones obtained within the first Born approximation~%
\cite{Matula_NJP16_053024:2014}.
We also consider the case when the twisted electron is a superposition
of two coherent vortex states.
For the second type of the target we investigate the dependence of the
experimentally measurable quantities on the position and size of the
target.
Besides, we present the analytical nonrelativistic expressions which 
allow one to check the results obtained by the numerical calculations
and to get a deeper insight into physics beyond them.
%
%
\\ \indent
%
%
The relativistic units ($m_e = \hbar = c = 1$) and the Heaviside charge
unit ($e^2 = 4\pi\alpha$) are used in the paper.
%
%
%
%
%
%
%
\section{BASIC FORMALISM}
The radiative recombination being the time-reversed photoionization
is the process in which a continuum electron is captured into
an ion bound state with the simultaneous emission of a photon.
The relativistic theory of the plane-wave electron RR is well established and vastly
presented in the literature (see, e.g., Refs.~\cite{Eichler,
Shabaev_PR356_119:2002, Surzhykov_JPB35_3713:2002, Eichler_PR439_1:2007}).
In the case of the twisted incident electron, only the nonrelativistic study
within the first Born approximation was performed~\cite{Matula_NJP16_053024:2014}.
Here we are focused on the systematic relativistic description of the twisted
electron RR with bare nuclei beyond the Born approximation.
Since the main aspects of this description are rather similar to 
those for the plane-wave case, we start with the brief recall of 
the plane-wave (conventional) electron RR theory.
%
%
\subsection{Radiative recombination of plane-wave (conventional) electrons}
%
%
The probability of the asymptotically plane-wave electron RR can be represented as follows
\begin{equation}
\frac{dW^{(\rm PW)}_{\bp\mu;m_f, \lambda}}{d\Omega_{\rm k}} =
2\pi\omega^2  \left| \tau^{(\rm PW)}_{\bp\mu;fm_f,\bk\lambda}  \right|^2,
\end{equation}
where $\bp$ and $\mu$ are the asymptotic momentum and the helicity of the incident
electron, respectively, $f$ denotes the final bound state, and the
emitted photon is characterized by the energy $\omega$, the momentum
$\bk$, and the polarization $\lambda$.
The amplitude of the RR process is given by
\begin{equation}
\tau^{(\rm PW)}_{\bp\mu;fm_f,\bk\lambda} =
\int d\br
\Psi^{\dagger}_{fm_f}(\br)
R_{\bk\lambda}^{\dagger}(\br)
\Psi^{(+)}_{\bp\mu}(\br),
\label{eq:ampl_pw}
\end{equation}
where $\Psi^{(+)}_{\bp\mu}$ and $\Psi_{fm_f}$ are the wave functions of the
electron in the initial and final states, respectively.
The transition operator in the Coulomb gauge has the following form
\begin{equation}
R_{\bk\lambda}(\br)
= -\sqrt{\frac{\alpha}{\omega(2\pi)^2}}\balpha\cdot\bepsilon_\lambda e^{i\bk\br}.
\end{equation}
Here $\balpha$ is the vector incorporating the Dirac matrices and 
$\bepsilon_\lambda$ is the photon polarization vector.
The wave function of the incident electron is constructed as the
solution of the Dirac equation in the external nucleus field with the
following asymptotic behaviour
\begin{equation}
\Psi^{(+)}_{\bp\mu}(\br)
\xrightarrow[r\rightarrow\infty]{}
\psi_{\bp\mu}(\br) + G_{\mu}^{(+)}(\bn_p,\bn)
\frac{e^{ipr}}{r}.
\label{eq:asym_pw_in}
\end{equation}
Here $\bn$ and $\bn_p$ are the unit vectors in the $\br$ and $\bp$
directions, respectively, $G^{(+)}$ is the bispinor amplitude, and the
plane-wave solution of the free Dirac equation expresses as
\begin{equation}
\psi_{\bp\mu}(\br) =
\frac{e^{i\bp\br}}{\sqrt{2\varepsilon(2\pi)^3}}u_{\bp\mu},
 \label{eq:psi_free}
\end{equation}
where $u_{\bp\mu}$ is the Dirac bispinor~\cite{Berestetsky, Eichler}
which satisfies the normalization condition $u^\dagger_{\bp\mu}u_{\bp\mu'} =
2\varepsilon\delta_{\mu\mu'}$. 
The explicit form of the wave function~(\ref{eq:asym_pw_in}) is given
by~\cite{Rose_RET, Pratt, Eichler}
\begin{equation}
\Psi^{(+)}_{\bp\mu}(\br) =
\frac{1}{\sqrt{4 \pi \varepsilon p}}
\sum_{\kappa m_{j}}
C^{j\mu}_{l0\ 1/2\mu}
i^{l}
\sqrt{2l + 1}
e^{i\delta_\kappa}
D^{j}_{m_{j}\mu}(\varphi_{p}, \theta_{p}, 0)
\Psi_{\varepsilon\kappa m_j} (\br),
\label{eq:wf_dir_in}
\end{equation}
where $\kappa = (-1)^{l + j + 1/2}(j + 1/2)$ is the Dirac quantum number
with $j$ and $l$ being the total and orbital angular momenta, respectively,
$C_{j_1m_1\ j_2m_2}^{JM}$ is the Clebsch-Gordan coefficient,
$\delta_\kappa$ is the phase shift being induced by the potential 
of the extended nucleus, $D_{MM'}^{J}$ is the Wigner
matrix~\cite{Rose_ETAM, Varshalovich}, and
$\Psi_{\varepsilon \kappa m_{j}} (\br)$ is the partial wave 
solution of the Dirac equation in the nucleus field~\cite{Berestetsky}.
Let us note here that at large distances the flux corresponding to 
the wave function~\eqref{eq:wf_dir_in} coincides with the flux of the 
free electron and equals
\begin{equation}
{\bf j}^{(\rm PW)} 
= 
\psi^{\dagger}_{\bp\mu}(\br)
\balpha
\psi_{\bp\mu}(\br)=
\frac{\bp}{(2\pi)^3 \varepsilon}.
\label{eq:flux_free}
\end{equation}
This fact is clearly seen from Eq.~\eqref{eq:asym_pw_in}.
The RR cross section reads
\begin{equation}
 \frac{d\sigma^{(\rm PW)}_{\bp\mu;m_f, \lambda}}{d\Omega_{\rm k}}=
 \frac{1}{|{\bf j}^{(\rm PW)}|}\,\frac{dW^{(\rm PW)}_{\bp\mu;m_f, \lambda}}{d\Omega_{\rm k}}.
 \label{eq:cross_section_pw}
\end{equation}
Here we would like to stress that in Eq.~\eqref{eq:cross_section_pw} 
the momentum direction of the incident electron is arbitrary with respect 
to the $z$ axis which is not yet fixed.
The presented formulas completely describe the process of the plane
wave RR.
%
%
\\
\indent
%
%
In what follows, we will often refer to the nonrelativistic theory of
the radiative recombination into the $1s$ state.
In this case, utilizing the dipole approximation one can obtain the
following formulas for the process probability and the cross 
section~\cite{Berestetsky}
\begin{equation}
\frac{dW^{(\rm PW,\ NR)}_{\lambda}}{d\Omega_{\rm k}} =
\frac{\alpha p}{(2\pi)^3}
\left| \bn_p \cdot \bepsilon_\lambda \right|^2
F(\nu),
\label{eq:prob_pw_nr}
\end{equation}
\begin{equation}
\frac{d\sigma^{(\rm PW,\ NR)}_{\lambda}}{d\Omega_{\rm k}} =
\alpha
\left| \bn_p \cdot \bepsilon_\lambda \right|^2
F(\nu),
\label{eq:dif_cross_nr}
\end{equation}
\begin{equation}
F(\nu) =
2^5\pi
\frac{\nu^6}{(1 + \nu^2)^2}
\frac{e^{-4\nu{\rm cot^{-1}}\nu}}{1-e^{-2\pi\nu}},
\label{eq:f_exact}
\end{equation}
where $p = |\bp|$ and $\nu = \alpha Z / p$.
In the Born approximation ($\nu \rightarrow 0$), the $F$
function~(\ref{eq:f_exact}) is given by
\begin{equation}
F_{B}(\nu) = 16\nu^5.
\label{eq:f_born}
\end{equation}
The corresponding nonrelativistic expressions for the RR into 
other states can be found in Refs.~\cite{Katkov, Milstein}.
%
%
\subsection{
Radiative recombination of asymptomatically twisted electrons
}
\begin{figure}[h!]
\centering
\includegraphics{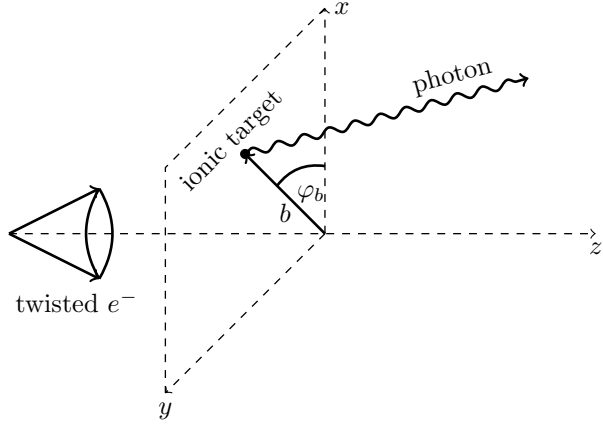}
\caption{The geometry of the RR process.}
\label{ris:geometry}
\end{figure}

%
%
Let us now switch to the description of the twisted electron RR.
As already been mentioned, a free twisted electron is characterized by
the following set of quantum numbers:
the energy $\varepsilon$, the helicity $\mu$, and the projections of the
momentum $p_z$ and the total angular momentum $m$ on the propagation 
direction. 
Here and throughout the $z$ axis is fixed along this direction.
Besides, the twisted electron possesses a well-defined absolute value of the transverse momentum $|\bp_\perp|\equiv \varkappa = \sqrt{\varepsilon^2 - 1 - p_z^2}$.
The vortex electron can be represented as a coherent superposition of
the plane waves with momenta forming the surface of a cone with the 
opening (conical) angle $\theta_p=\arctan(\varkappa/p_z)$.
The explicit expression for the wave function of the free twisted
electron is given by~\cite{Serbo_PRA92_012705:2015}
\begin{equation}
\psi_{\varkappa m p_{z} \mu}(\br) =
\int
d\bp
\frac{e^{im\varphi_{p}} }{2\pi p_\perp}
\delta(p_{\parallel} - p_{z})
\delta(p_{\perp} - \varkappa)
i^{\mu - m}
\psi_{\bp\mu}(\br),
\label{eq:tw_free}
\end{equation}
where $p_\parallel$ and $p_\perp$ are the longitudinal and perpendicular 
components of momentum $\bp$, respectively.
In the plane-wave limit, this wave function behaves as
\begin{equation}
\psi_{\varkappa m p_z \mu}(\br)
\xrightarrow[\theta_p \rightarrow 0]{}
\delta_{\mu m}
\psi_{\tilde{\bp}\mu}(\br), \;\;
\tilde{\bp}=(0,0, p_z).
\label{eq:tw_free_parax}
\end{equation}
From Eq.~(\ref{eq:tw_free}) it is seen that the density and
the flux of the twisted electron are not the homogeneous functions of
the space variables. In particular, the density equals
\begin{equation}
\rho_{m\mu}^{(\rm tw)}(\br_\perp) =
\psi^{\dagger}_{\varkappa m p_z \mu}(\br)\psi_{\varkappa m p_z \mu}(\br)=
\frac{1}{(2\pi)^3}
\sum_{\sigma}\left[d^{\,\,1/2}_{\sigma\mu}(\theta_p)\right]^2\, 
J^2_{m-\sigma}(\varkappa r_\perp),
\label{eq:density_free}
\end{equation}
where $\br_\perp$ is the perpendicular component of $\br$ and 
$r_\perp = |\br_\perp|$.
Therefore, in contrast to the plane wave case the relative position of
the twisted electron and the target is important.
For the target ion being shifted from the $z$ axis on the impact
parameter $\bb$ (see Fig.~\ref{ris:geometry}) the amplitude of the RR
process is given by
\begin{equation}
\tau^{(\rm tw)}_{\varkappa m p_z\mu;fm_f,\bk\lambda}(\bb) =
\int d\br
\Psi^{\dagger}_{fm_f}(\br - \bb)
R_{\bk\lambda}^{\dagger}(\br)
\Psi^{(+)}_{\varkappa m p_{z} \mu}(\br),
\label{eq:amp_tw}
\end{equation}
where $\Psi^{(+)}_{\varkappa m p_{z} \mu}(\br)$ is the wave function of the
twisted electron.
For practical calculations it is more convenient to change the
integration variable in Eq.~(\ref{eq:amp_tw}) as follows
$\br - \bb \rightarrow \br$.
For such the geometry, the wave function of the twisted electron
is to be taken as the solution of the Dirac equation in the central
field with the following asymptotics~\cite{Schweber}
\begin{equation}
\Psi^{(+)}_{\varkappa m p_{z} \mu}(\br + \bb)
\xrightarrow[r\rightarrow\infty]{}
\psi_{\varkappa m p_{z} \mu}(\br + \bb) +
G^{(\rm tw)}_{m\mu,\bb}(\theta_p,\bn) \frac{e^{ipr}}{r}.
\label{eq:wf_tw_asymp}
\end{equation}
The corresponding solution is given by
\begin{eqnarray}
\Psi^{(+)}_{\varkappa m p_{z} \mu}(\br + \bb)
& = &
\int
d\bp
\frac{e^{im\varphi_{p}} }{2\pi p_{\perp}}
\delta(p_{\parallel} - p_{z})
\delta(p_{\perp} - \varkappa)
i^{\mu - m}
e^{i\bp\cdot\bb}
\Psi^{(+)}_{\bp \mu}(\br)
\label{eq:wf_tw_imp1}
\\ \nonumber
& = &
\frac{1}{\sqrt{4\pi\varepsilon p}}
\sum_{\kappa m_j}
i^{l + \mu - m_j}
C^{j\mu}_{l0\ 1/2\mu}
\sqrt{2l+1}
e^{i\delta_{\kappa}}
e^{-im_j\varphi_b}
d_{m_j\mu}^{j}\left(\theta_p\right)
\Psi_{\varepsilon\kappa m_j}(\br)
\\ && \times
e^{im\varphi_b}J_{m-m_j}(\varkappa b).
\label{eq:wf_tw_imp2}
\end{eqnarray}
Utilizing Eq.~(\ref{eq:wf_tw_imp1}) one can obtain the following
expression for the amplitude of the process under consideration:
\begin{equation}
\tau^{(\rm tw)}_{\varkappa m p_z\mu;fm_f,\bk\lambda}(\bb)
=
e^{-i\bk\cdot\bb}
\int
\frac{e^{im\varphi_{p}} }{2\pi p_{\perp}}
\delta(p_{\parallel} - p_{z})
\delta(p_{\perp} - \varkappa)
i^{\mu - m}
e^{i\bp\cdot\bb}
\tau^{(\rm PW)}_{\bp\mu;fm_f,\bk\lambda}
d\bp.
\label{eq:ampl_tw}
\end{equation}
Then the probability of the twisted electron RR is given by
\begin{equation}
\frac{dW^{(\rm tw)}_{m\mu;m_f,\lambda}}{d\Omega_k}(\bb)
=
2\pi\omega^2  \left|
\int
d\bp
\frac{e^{im\varphi_{p}} }{2\pi p_{\perp}}
\delta(p_{\parallel} - p_{z})
\delta(p_{\perp} - \varkappa)
i^{\mu - m}
e^{i\bp\cdot\bb}
\tau^{(\rm PW)}_{\bp\mu;fm_f,\bk\lambda}
\right|^2.
\label{eq:prob_sngl_ion_p}
\end{equation}
%
%
\\
\indent
%
%
Since all measurable quantities can be expressed in terms of the probability~%
(\ref{eq:prob_sngl_ion_p}) we regard the theoretical description of the
twisted electron RR as completed.
Here it should be emphasized that the wave function, which is introduced in
Eqs.~(\ref{eq:wf_tw_imp1}) and~(\ref{eq:wf_tw_imp2}), accounts for the
interaction of the asymptotically twisted electron with the
target ion in all orders in $\alpha Z$.
Thus, utilizing this wave function one obtains the results beyond the Born
approximation.
%
%
\subsection{Measurables}
%
%
Presently, the experiments with a single ion, especially heavy and
highly-charged one, are very difficult and time consuming.
Therefore, in the present paper, we focus on the analysis of the
twisted electron RR with various targets.
The distribution of the ions within the target can be considered as the 
classical one and is assumed to be given by the function $f(\br_{\perp})$
with the following normalization condition
\begin{equation}
\int d\br_{\perp} f(\br_{\perp}) = 1.
\end{equation}
In this case, all measurables are determined via the integral of this
function with the probability of the RR with the ion located at the
distance $\bb$ from the $z$ axis
\begin{equation}
\frac{d\overline{W}^{(\rm tw)}_{m\mu;m_f,\lambda}}{d\Omega_k}(\bb_t)
=
\int d\bb f(\bb - \bb_t)
\frac{dW^{(\rm tw)}_{m\mu;m_f,\lambda}}{d\Omega_k}(\bb).
\label{eq:prob_mes}
\end{equation}
Here $\bb_t$ stands for the coordinates of the target centre.
%
%
\\
\indent
%
%
One of the main process characteristics, the cross section, cannot be
determined for the twisted electrons as a ratio of the probability to
the flux density of the incoming particles.
Indeed, it is clear from the free twisted electron wave function~%
(\ref{eq:tw_free}) that, in contrast to the plane wave case,
the flux is neither a homogeneous function nor even positively defined.
Nevertheless, it is very useful to have an ``effectively''
defined cross section.
For example, it can be used for the estimation of the experimental
feasibility.
In the present paper, we propose the following expression for the
cross section
\begin{equation}
\frac{d\sigma^{(\rm tw)}_{m\mu;m_f,\lambda}}{d\Omega_k}(\bb_t)
=
\frac{1}{J_z}
\frac{d\overline{W}^{(\rm tw)}_{m\mu;m_f,\lambda}}{d\Omega_k}(\bb_t),
\label{eq:cross_serbo}
\end{equation}
\begin{equation}
J_{z} =
v_z\int d\br_\perp f(\br_\perp)\rho_{1/2}^{(\rm tw)}(\br_\perp),
\label{eq:flow_serbo}
\end{equation}
where $v_z = (p/\varepsilon)\,\cos\theta_p $ and 
$\rho_{1/2}^{(\rm tw)}(\br_\perp)$ is the density of the free twisted 
electron~\eqref{eq:density_free} with $m = \mu = 1/2$.
Both the cross section~(\ref{eq:cross_serbo}) and the
flux~(\ref{eq:flow_serbo}) goes to the well known conventional
expressions~\eqref{eq:cross_section_pw} and \eqref{eq:flux_free}, respectively, in the 
plane-wave (paraxial) limit.
This fact is regarded as the main argument in favour of the
definitions~\eqref{eq:cross_serbo}-\eqref{eq:flow_serbo}.
However, it is worth noting that one can easily present a set of
different cross section determinations possessing the same limit.
%
%
\\
\indent
%
%
In addition to the cross section, one can characterize the twisted
electron RR by the relative measurable quantities.
One of them is normalized on average angular probability
\begin{equation}
\frac{d\overline{W}_{\rm norm}}{d\Omega_k}(\bb_t)
=
\frac{1}{\overline{W}^{\rm (avr)}_{m}(\bb_t)}
\frac{1}{2}
\sum_{\mu m_f\lambda}
\frac{d\overline{W}_{m\mu;m_f,\lambda}}{d\Omega_k}(\bb_t),
\label{eq:ang_norm}
\end{equation}
\begin{equation}
\overline{W}^{\rm (avr)}_{m}(\bb_t)
=
\frac{1}{4\pi}
\int d\Omega_k
\frac{1}{2}
\sum_{\mu m_f\lambda}
\frac{d\overline{W}_{m\mu;m_f,\lambda}}{d\Omega_k}(\bb_t).
\end{equation}
The relative variables also include the Stokes parameters
\begin{equation}
P_{l} = P_1
= \frac{W_{0^\circ} - W_{90^\circ}}{W_{0^\circ} + W_{90^\circ}},
\qquad
P_{2}
= \frac{W_{45^\circ} - W_{135^\circ}}{W_{45^\circ} + W_{135^\circ}},
\qquad
P_{3} = P_{c}
= \frac{W_{+1} - W_{-1}}{W_{+1} + W_{-1}}.
\label{eq:stokes}
\end{equation}
Here $W_{\lambda}$ denotes $\frac{1}{2}\sum_{\mu m_f}
\frac{d\overline{W}_{m\mu;m_f,\lambda}}{d\Omega_k}(\bb_t)$
meanwhile $W_{\chi}$ designates the probability of the photon emission
with the linear polarization $\bepsilon_\chi = \frac{1}{\sqrt{2}}
\sum_{\lambda = \pm 1}e^{-i\lambda\chi}\bepsilon_\lambda$.
%
%
%
%
%
%
%
%
\section{RESULTS AND DISCUSSIONS}
The radial parts of the bound- and continuum-state wave functions being 
the solutions of the Dirac equation in the central field of the extended 
nucleus are numerically found utilizing the modified RADIAL package~%
\cite{Salvat_CPC90_151:1995}. 
The Fermi model of the nuclear charge distribution is employed.
In order to reach the convergence of the results the partial waves with 
$|\kappa|$ up to $10$ are taken into account.
%
%
\\
\indent
%
%
There are two distinct types of experiments.
In the first one, the target has a macroscopic size and therefore can be
regarded as infinite.
A target, which consists of a finite number of ions (up to a single ion),
forms the second type and is referred to as the mesoscopic one.
In order to describe the ion distribution for both types of the
targets we choose the Gaussian distribution which is generally
realized in ion traps. 
Then, the function $f$ reads
\begin{equation}
f(\bb - \bb_t) = \frac{1}{2\pi w^2}
e^{-\frac{(\bb - \bb_t)^2}{2w^2}},
\label{eq:ion_distr}
\end{equation}
where $\bb_t$ corresponds to the centre of the target and the dispersion 
$w$ characterizes the size of the target.
The macroscopic target corresponds to the limit $w \rightarrow \infty$,
while at $w \rightarrow 0$ one obtains the case of a single ion.
For this distribution, the flux defined by Eq.~(\ref{eq:flow_serbo})
takes the following form
\begin{equation}
J_{z}
=
e^{-(w\varkappa)^2}\frac{p\cos\theta_p}{\varepsilon (2\pi)^3}
\left[
\cos^2\frac{\theta_p}{2}
I_{0}(\varkappa^2 w^2)
+
\sin^2\frac{\theta_p}{2}
I_{1}(\varkappa^2 w^2)
\right],
\label{eq:flow_serbo_app}
\end{equation}
where $I_{n}$ is the modified Bessel function of the first kind~\cite{Watson, Abramowitz}.
%
%
\subsection{Macroscopic target}
%
%
The macroscopic target is the simplest one for the experimental
realization as well as for the theoretical investigation.
As it was mentioned above, such a target can be described by the
function~(\ref{eq:ion_distr}) with $w \rightarrow \infty$.
This corresponds to the infinite spatial size with the uniform
distribution of ions inside the target.
With this in mind, one can utilize $f = 1/(\pi R^2)$ with $R = 2w\sqrt{2/\pi}$
(the radius of the cylindrical box) instead of the Gaussian
distribution~(\ref{eq:ion_distr}).
Repeating the calculations of Ref.~\cite{Serbo_PRA92_012705:2015} we obtain
the differential cross section for the macroscopic target in the simple form
\begin{equation}
\frac{d\sigma^{(\rm mac)}_{\mu; m_f, \lambda}}{d\Omega_k}
=
\frac{1}{\cos\theta_p}
\int_{0}^{2\pi} \frac{d\varphi_p}{2\pi}
\frac{d\sigma^{(\rm PW)}_{\bp\mu; m_f, \lambda}}{d\Omega_k},
\label{eq:cross_mac}
\end{equation}
where $d\sigma^{(\rm PW)}_{\bp\mu; m_f, \lambda}/d\Omega_k$ is defined
by Eq.~\eqref{eq:cross_section_pw}. 
Note, that this cross section is $m$ and $\bb_t$ independent.
In addition, from Eq.~\eqref{eq:cross_mac} one can obtain the following 
relation for the total cross section being averaged over $\mu$ and 
summed over $m_f$ and $\lambda$
\begin{equation}
\sigma^{(\rm mac)}_{\rm tot} 
= \frac{\sigma^{(\rm PW)}_{\rm tot}}{\cos\theta_p}.
\label{eq:cross_tw_tot}
\end{equation}
%
%
%
\subsubsection{
Comparison of the Born approximation with the exact treatment
}
\begin{figure}[h!]
\centering
\includegraphics{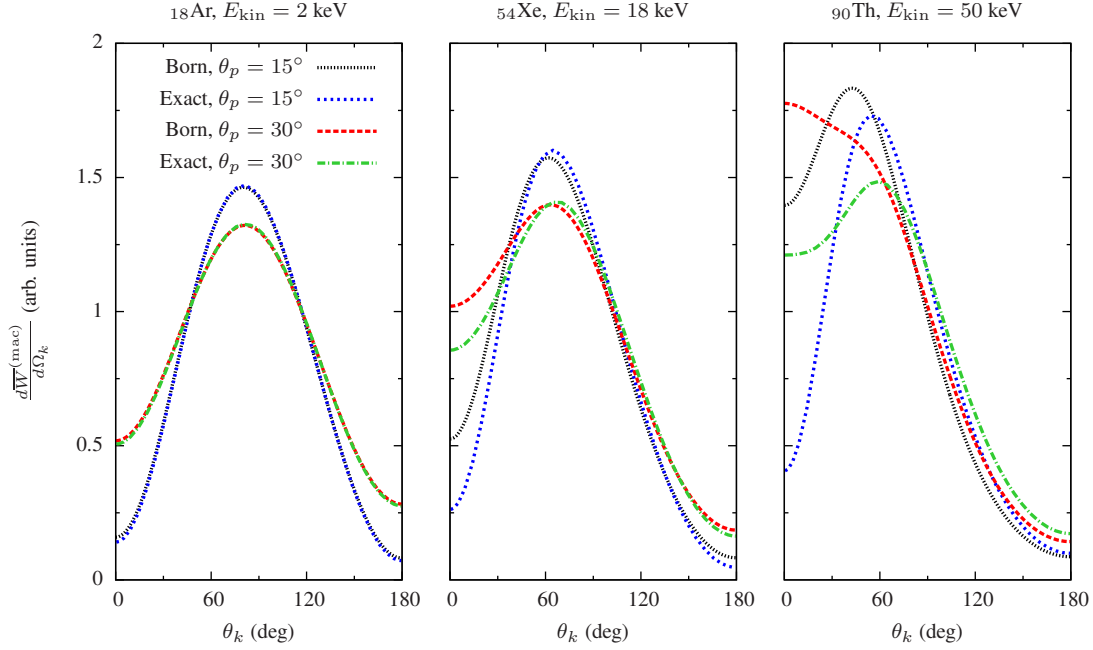}
\caption{
The normalized angular distribution~(\ref{eq:ang_norm}) for the RR of the
twisted electron into the $1s$ state of H-like ions.
On the left, middle, and right panels the cases of the argon ($Z = 18$),
xenon ($Z = 54$), and thorium ($Z = 90$) ions are presented, respectively.
The kinetic energy of the incident electron is
$2$ keV (for argon), $18$ keV (for xenon), and $50$ keV (for thorium).
}
\label{ris:macr_1s}
\end{figure}

\begin{figure}[h!]
\centering
\includegraphics{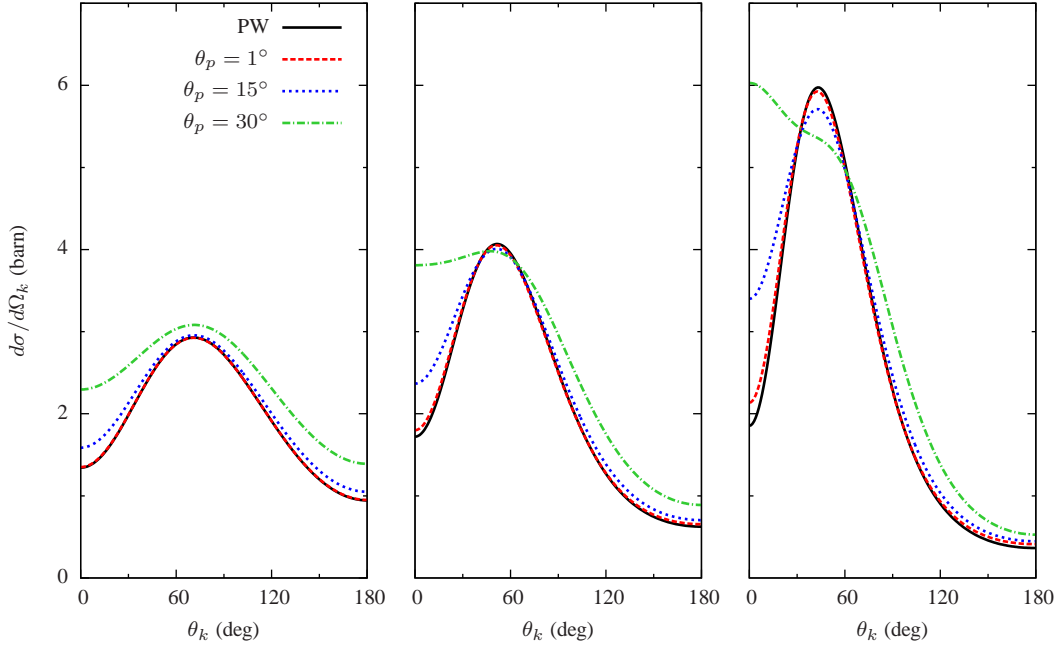}
\caption{
The differential cross section~(\ref{eq:cross_mac}) for the RR of the
twisted electron into the $2p_{3/2}$ state of H-like ions.
On the left, middle, and right panels the cases of the argon ($Z = 18$),
xenon ($Z = 54$), and thorium ($Z = 90$) ions are presented, respectively.
The kinetic energy of the incident electron is
$2$ keV (for argon), $18$ keV (for xenon), and $50$ keV (for thorium).
}
\label{ris:macr_2p32}
\end{figure}

\begin{figure}[h!]
\centering
\includegraphics{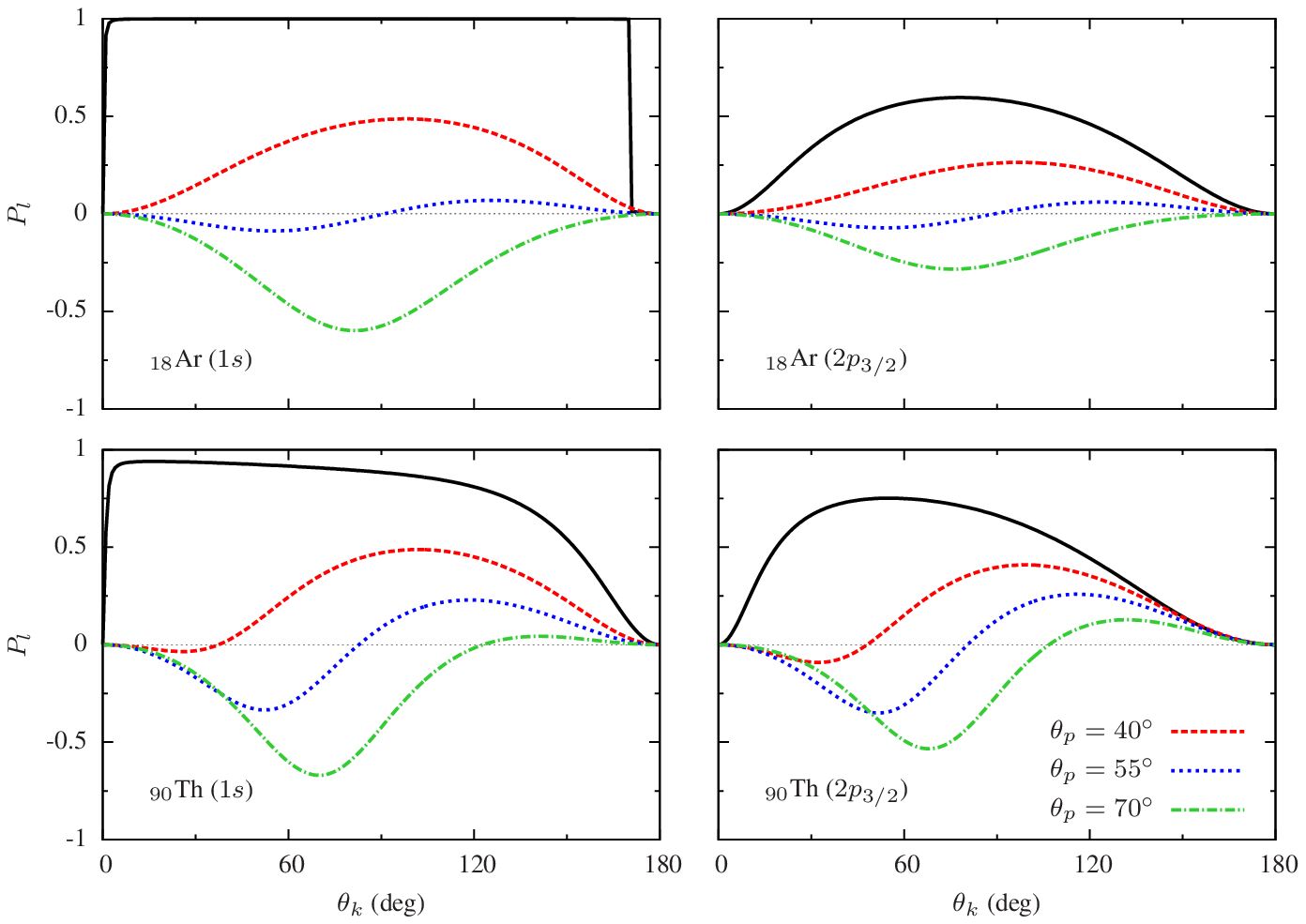}
\caption{
The degree of the linear polarization~(\ref{eq:stokes}) for the RR of the
twisted electron with the bare nuclei.
The results for the argon ($Z = 18$) ion at $2$ keV electron energy are
presented in the first row.
In the second row, the case of the thorium ($Z = 90$) ion at the 
$50$ keV electron energy is depicted.
The recombinations into the $1s$ and $2p_{3/2}$ states are 
presented in the left and right columns , respectively.
The black solid line corresponds to the conventional plane-wave 
asymptotics case.
}
\label{ris:macr_pol}
\end{figure}

%
%
Let us first consider the RR into the $1s$ state of a H-like ion.
In this case, the exact nonrelativistic expression for the differential
cross section is given by Eq.~\eqref{eq:dif_cross_nr}.
In order to investigate the importance of the calculations beyond the
Born approximation we introduce the following parameter
\begin{equation}
R_{\rm NR}(\nu) = \frac{d\sigma^{(\rm NR)}_{\lambda} / d\Omega_k}
              {d\sigma^{(\rm NR,\ B)}_{\lambda} / d\Omega_k}
=
\frac{2\pi\nu}{(1+\nu^2)^2}
\frac{e^{-4\nu {\rm arcctg}\nu}}{1 - e^{-2\pi\nu}}.
\label{eq:ratio}
\end{equation}
From Eq.~(\ref{eq:cross_mac}), it is clearly seen that the $R_{\rm NR}$
parameter takes the same values for both the plane-wave and twisted electrons.
Additionally, one can conclude that the ratio~(\ref{eq:ratio}) does not depend on
the parameters of the outgoing photon.
It equals to $1$ at $\nu = 0$ and rapidly decreases with the growth of the
$\nu$ parameter.
For the process discussed in Ref.~\cite{Matula_NJP16_053024:2014}, where
the $2$ keV twisted electron RR into the $1s$ state of the hydrogen ion
($\nu = 0.083$) was studied, one gets $R_{\rm NR} = 0.77$.
This corresponds to the 23\% difference between the results obtained within the Born approximation and beyond it.
In the case of the recombination with the argon ($Z = 18$) 
ion at the same electron energy $R_{\rm NR} = 0.03$!
This means that the Born approximation does not provide reliable results 
for the absolute value of the differential cross section.
%
%
\\
\indent
%
%
The situation differs for the relative values of the
measurables~(\ref{eq:ang_norm}) and (\ref{eq:stokes}).
The explicit nonrelativistic expression for the angular distribution is
\begin{equation}
\frac{d\overline{W}_{\rm norm}^{(\rm tw,\ NR)}}{d\Omega_{k}}
= \frac{3}{4}
\left[
\left(2 - 3\sin^2\theta_p\right)\sin^2\theta_{k}
+ 2\sin^2\theta_p
\right],
\label{eq:prob_tw_nr}
\end{equation}
and the Stokes parameters are given by
\begin{equation}
P_{l}^{(\rm tw,\ NR)}
=
\frac{\left(2-3\sin^2\theta_p\right)\,\sin^2\theta_k}
 {\left(2-3\sin^2\theta_p\right)\,\sin^2\theta_k+
 2\sin^2\theta_p},
\qquad
P_2^{(\rm tw,\ NR)}  = P_c^{(\rm tw,\ NR)}  = 0.
\label{eq:stokes_tw_nr}
\end{equation}
Here we have substituted Eq.~(\ref{eq:prob_pw_nr}) into
Eq.~(\ref{eq:cross_mac}) and utilized the relation
\begin{equation}
\int \frac{d\varphi_{p}}{2\pi} \left|\bn_p {\bf e}\right|^2
=
\frac{1}{2}\left[\left(2-3\sin^2\theta_p\right)\,|e_z|^2+
 \sin^2\theta_p\right],
\end{equation}
where $\bf e$ is an arbitrary unit vector.
The corresponding expressions for the conventional case can be obtained by
letting $\theta_p \rightarrow 0$.
As an example, the degree of linear polarization $P_{l}^{(\rm PW,\ NR)} 
= 1$.
From Eq.~\eqref{eq:prob_tw_nr} one can see that the
angular distributions being calculated within and beyond the Born
approximation coincide with each other.
The same is valid for the Stokes parameters~(\ref{eq:stokes_tw_nr}).
The results obtained by the usage of Eqs.~(\ref{eq:prob_tw_nr})
and (\ref{eq:stokes_tw_nr}) are in excellent agreement (up to the terms 
of order $\omega/p\ll 1$) with the ones presented in 
Ref.~\cite{Matula_NJP16_053024:2014}.
%
%
\\
\indent
%
%
Here it is worth stressing that the coincidence of the relative 
measurable values being
calculated within and beyond the Born approximation occurs only
in the nonrelativistic framework.
This is not the case in the relativistic formalism.
In Fig.~\ref{ris:macr_1s} we present the normalized angular 
distribution for the RR of the twisted electron into the $1s$ state.
The kinetic energies $E_{\rm kin}$ were chosen to
provide the same parameter $\nu = 1.48$ ($R_{\rm NR} = 0.03$) for 
all the ions.
Fig.~\ref{ris:macr_1s} demonstrates the difference between the results obtained 
with the usage of the Born approximation and beyond it.
The comparison indicates the importance of the exact relativistic calculations for 
the systems with middle and high $Z$.
Indeed, for the uranium ion at $\theta_p = 30^\circ$ there is a 
qualitative difference in the behaviour of the differential cross sections.
Specifically, the forward photon emission becomes preferable in this case.
In Fig.~\ref{ris:macr_2p32} we present the differential cross section
for the RR of the twisted electron into the $2p_{3/2}$ state.
From this figure one can see that the role of the electron 
twistedness increases with the growth of $Z$.
%
%
\\
\indent
%
%
Let us now consider the outgoing photon polarization.
For an initially plane-wave electron, the degree of the linear polarization 
$P_{l}$ takes only positive values.
In the case of the twisted electron, the $P_{l}$ Stokes parameter becomes 
negative at $\theta_p> \arcsin\sqrt{2/3}\approx 55^\circ$ (see 
Eq.~\eqref{eq:stokes_tw_nr}).
This means that the emitted photon is linearly polarized in the
direction perpendicular to the scattering plane.
A similar effect has been observed in Ref.~\cite{Ivanov_PRA93_053825:2016}
where the Vavilov-Cherenkov radiation by twisted electrons has been
studied.
The $P_{l}$ Stokes parameter, which was calculated using the 
relativistic formalism beyond the Born approximation, is presented in 
Fig.~\ref{ris:macr_pol}.
From this figure one can see that in the case of the argon ($Z = 18$)
ion the photon polarization becomes negative at $\theta_p \sim 55^\circ$.
This is in a good agreement with the predictions by Eq.~(\ref{eq:stokes_tw_nr}).
For the much heavier thorium ($Z = 90$) ion $P_l$ changes its sign
already at $\theta_p \sim 40^\circ$.
Such a shift to smaller conical angles at higher $Z$ is due to the more
pronounced manifestation of the electron twistedness.
%
%
\subsubsection{
The RR of the electron being in a superposition of two vortex states
}
\begin{figure}[h!]
\centering
\includegraphics{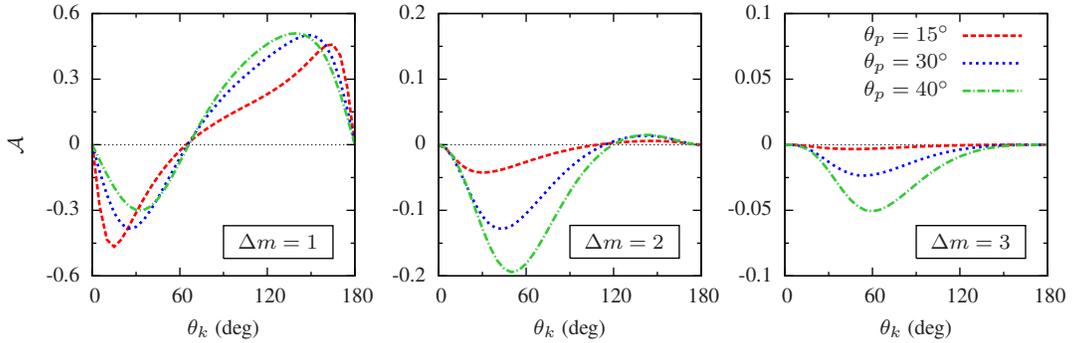}
\caption{
The azimuthal asymmetry parameter $\mathcal{A}$, defined by 
Eq.~\eqref{eq:cross_asym}, for the radiative recombination of the $18$ keV 
twisted electron into the $1s$ state of the xenon ($Z = 54$) ion.
It is assumed that $|c_1c_2| = 1/2$ and $\Delta\alpha = 0$.
}
\label{ris:superposition}
\end{figure}

%
%
It is of special interest the situation when the twisted
electron is not an eigenstate of the $J_z$ operator but a coherent
superposition of such states.
As an example, let the superposition consists of two twisted waves with
different $m$~\cite{Ivanov_PRD83_093001:2011}.
In order to obtain the wave function of such an incident electron 
one has to perform the following substitution in Eq.~(\ref{eq:wf_tw_imp1})
\begin{equation}
i^{-m}e^{im\varphi_p} \rightarrow c_1 i^{-m_1} e^{im_1\varphi_p}
                                + c_2 i^{-m_2} e^{im_2\varphi_p},
\end{equation}
where the complex coefficients $c_n = |c_n|e^{i\alpha_n}$ satisfy the
normalization condition $|c_1|^2 + |c_2|^2 = 1$.
As a result of this substitution the differential cross section~%
(\ref{eq:cross_mac}) takes the form
\begin{equation}
\frac{d\sigma_{\mu;m_f,\lambda}^{(\rm sup)}}{d\Omega_k}
=
\frac{1}{\cos\theta_p}
\int \frac{d\varphi_p}{2\pi}G(\varphi_p)
\frac{d\sigma_{\bp\mu;m_f,\lambda}^{(\rm PW)}}{d\Omega_k},
\label{eq:cross_sup}
\end{equation}
where
\begin{equation}
G(\varphi_p) = 1
+ 2|c_1c_2|
\cos\left[\Delta m\left(\varphi_p - \pi/2\right) + \Delta\alpha\right],
\quad
\Delta m = m_2 - m_1,
\quad
\Delta\alpha = \alpha_2 - \alpha_1.
\label{eq:g_factor}
\end{equation}
In Ref.~\cite{Matula_NJP16_053024:2014}, it was pointed out that the
presence of this additional $G$ factor leads to a modification of the
angular distribution and the Stokes parameters $P_l$ and $P_2$.
Here we will focus only on the modification of the differential cross
section.
It can be shown that after the summation and averaging over the final and 
initial states projections, respectively, the differential cross 
section~\eqref{eq:cross_sup} can be written in the following form
\begin{equation}
\frac{d\sigma^{(\rm sup)}}{d\Omega_k}
=
\frac{d\sigma^{(\rm mac)}}{d\Omega_k}
 \left\{ 1+
\mathcal{A}
\cos\left[\Delta m \left(\varphi_k - \pi/2\right) + \Delta\alpha\right]
\right\},
\label{eq:cross_asym}
\end{equation}
where $d\sigma^{(\rm mac)}/d\Omega_k$ is the differential cross 
section~\eqref{eq:cross_mac} being averaged over $\mu$ and summed over 
$m_f$ and $\lambda$.
In the nonrelativistic case, substituting Eq.~\eqref{eq:prob_pw_nr} into 
Eq.~\eqref{eq:cross_sup} and summing over the polarization of the 
emitted photon one can obtain the explicit expression for the azimuthal 
asymmetry parameter
\begin{equation}
\mathcal{A}_{\rm NR} = -\frac{|c_1 c_2|}{\left(2-3\sin^2\theta_p\right) \sin^2\theta_k  +2 \sin^2\theta_p}\cdot \left\lbrace
\begin{aligned}
& \sin2\theta_k\sin2\theta_p && {\rm at}\ \Delta m = \pm 1
\\
& \sin^2\theta_k\sin^2\theta_p && {\rm at}\ \Delta m = \pm 2
\\
& 0 && {\rm otherwise}.
\end{aligned}
\right.
\end{equation}
From this expression it is clearly seen that the differential 
cross section possesses the azimuthal asymmetry only at $\Delta m = \pm 1$ 
or $\pm 2$ (we assume that $\Delta m \neq 0$).
Here it is worth mentioning that these selection rules originate from the
dipole approximation which was used to derive Eq.~\eqref{eq:prob_pw_nr}.
However, these rules partly take place in the exact relativistic calculations too.
This can be explained as follows.
The azimuthal asymmetry appears due to the interference of the RR
amplitudes being related to different partial waves
$\Psi_{\varepsilon\kappa m_j}$ in the decomposition~(\ref{eq:wf_tw_imp2}).
The higher $\Delta m$, the higher $\kappa$ are required.
The partial amplitudes decrease with the growth of $\kappa$ that leads
to a decrease of the asymmetry.
As a result, the manifestation of the asymmetry is more prominent at
$\Delta m = \pm 1$ and less at $\Delta m = \pm 2$.
In Fig.~\ref{ris:superposition}, the azimuthal asymmetry parameter 
$\mathcal{A}$ being obtained within the relativistic framework is 
depicted.
From this figure it is seen that the asymmetry is the most pronounced at
$\Delta m = \pm 1$. 
Nevertheless, the $\mathcal{A}$ parameters for $\Delta m = \pm 1$ and 
$\Delta m = \pm 2$ become comparable with each other at large conical angles $\theta_p$.
%
%
%
\subsection{Mesoscopic target}
\begin{figure}[h!]
\centering
\includegraphics{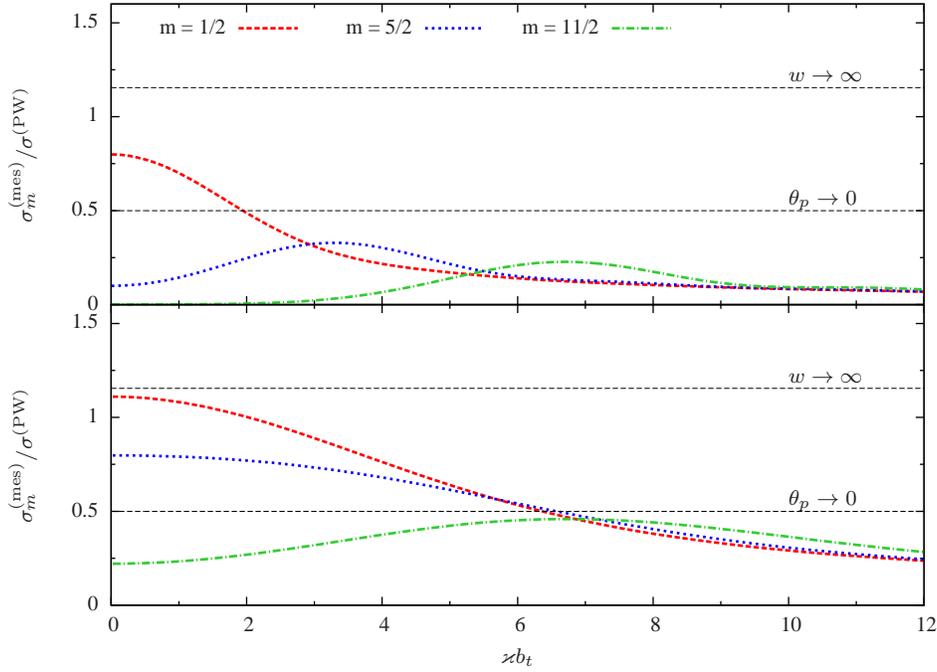}
\caption{
The total cross section of the $1$ keV twisted electron RR into the
$1s$ state of the argon ($Z = 18$) ion.
It is assumed that $\theta_p = 30^\circ$.
The cases $w = 1/\varkappa$ and $w = 3/\varkappa$ are
presented in the upper and lower graphs, respectively.
The limits $w\rightarrow \infty$ and $\theta_p\rightarrow 0$ are 
also displayed.
}
\label{ris:mes_tot}
\end{figure}

\begin{figure}[h!]
\centering
\includegraphics{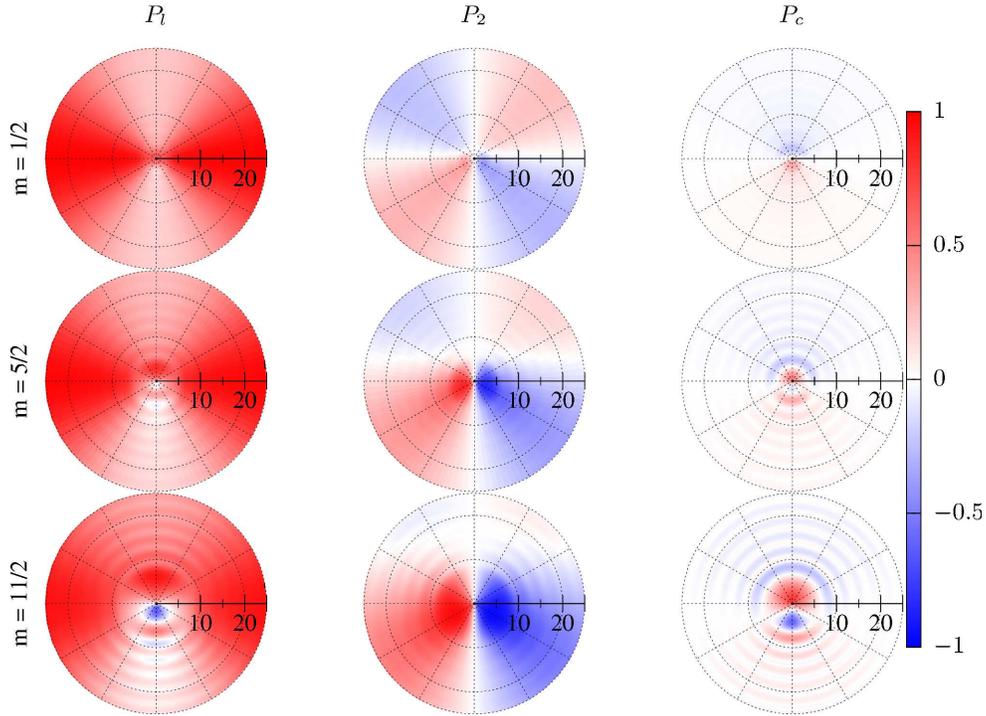}
\caption{
The Stokes parameters~(\ref{eq:stokes}) for the $1$ keV twisted electron
RR into $1s$ state of the argon ($Z = 18$) ion as a functions of the
target position.
The distance to the target is represented in terms of the dimensionless
variable $\varkappa b_t$.
It is assumed that $\theta_p = 30^\circ$, $\theta_k = 45^\circ$, and
$w = 1/\varkappa$.
}
\label{ris:mes_pol}
\end{figure}

%
%
Let us now consider the targets of the limited size.
In this case, the measurables appear to be sensitive to the total angular 
momentum projection $m$ on the propagation direction.
These targets are also sensitive to the spatial structure of the
incoming vortex particles~\cite{Peshkov_PS91_064001:2016,
Peshkov_PRA92_043415:2015, Schmiegelow_EPJD66_1:2012, Schmiegelow_arxiv}.
Here we present the results only for the case of the bare argon
($Z = 18$) nucleus.
Mesoscopic target consisting of such ions can be, in principle, created
nowadays~\cite{Schmoger_S347_1233:2015}.
%
%
\\
\indent
%
%
Let us start from the consideration of the total cross section
\begin{equation}
\sigma^{(\rm mes)}_{m, {\rm tot}}(b_t) =
\frac{1}{2}\sum_{\mu} \sum_{m_f \lambda}
\int d\Omega_k \frac{d\sigma^{(\rm tw)}_{m\mu;m_f,\lambda}}{d\Omega_k}(\bb_t),
\label{eq:mes_cross_tot}
\end{equation}
where $b_t$ is the target position and $d\sigma^{(\rm tw)}_{m\mu;m_f,\lambda} /
d\Omega_k$ is defined by Eq.~(\ref{eq:cross_serbo}) with the flux being
given by Eq.~(\ref{eq:flow_serbo_app}).
From Eq.~(\ref{eq:mes_cross_tot}) it can be seen that the total cross
section is independent of $\varphi_b$.
The ratio of this cross section to the plane-wave one is depicted in
Fig.~\ref{ris:mes_tot} as a function of the target position.
From this figure it is seen that for $w = 3/\varkappa$ the total cross
section appears to be less sensitive to $m$ and, as a result, to the
spatial structure of the incoming electron state.
Therefore, in what follows we will consider only the case $w = 1/
\varkappa$.
At $w \rightarrow \infty$ the ratio which is presented in 
Fig.~\ref{ris:mes_tot} goes to $1/\cos\theta_p$ that corresponds to the case of a macroscopic target~(see Eq.~\eqref{eq:cross_tw_tot}).
In addition, the $m$ dependence becomes much less pronounced.
The situation changes at $w$ fixed and $\theta_p \rightarrow 0$.
In this case, the ratio $\sigma_{m}^{(\rm mes)} / \sigma^{(\rm PW)}$
equals to $1/2$ at $m = 1/2$ and zero at $m \neq 1/2$.
This can be explained as follows.
At $\theta_p \rightarrow 0$ the transverse momentum $\varkappa
\rightarrow 0$.
Therefore, the projection of the total angular momentum on the propagation
direction equals to the spin projection on the momentum ($\mu = m$).
As a result, in the averaging over the helicities ($\frac{1}{2}\sum_{\mu}$)
in Eq.~(\ref{eq:mes_cross_tot}) only one term with $\mu = m$ contributes
and the $1/2$ factor remains.
The ratio which is depicted in Fig.~\ref{ris:mes_tot} goes exactly to
the $1/2$ factor at the limit $\theta_p \rightarrow 0$.
In order to get the ``correct'' paraxial limit, namely $\sigma_{m}^{(\rm mes)}
/ \sigma^{(\rm PW)} \rightarrow 1$, one has to put simultaneously
$\theta_p \rightarrow 0$ and $w\theta_p \rightarrow \infty$.
%
%
\\
\indent
%
%
The Stokes parameters are depicted in Fig.~%
\ref{ris:mes_pol} as functions of the target position for different
$m$ values.
From this figure one can observe a strong correlation between the target
position and the polarization of the emitted photon.
It can also be seen that the correlation increases with the growth of $m$.
Thus, one can investigate the spatial structure of the
twisted electron via measuring the Stokes parameters of the RR photon
for different target positions.
Alternatively, the target position can be determined by studying the
polarization of the emitted radiation.
%
%
\\
\indent
%
%
Here it is worth mentioning that $w = 1 / \varkappa$ for the $1$ keV
twisted electron with $\theta_p = 30^\circ$ corresponds to the target
size about $0.01$ nm.
Therefore, one needs to utilize focused twisted electron beams of a 
sub-nanometer size.
The possibility of generating such beams was demostrated in Refs.~%
\cite{Verbeeck_APL99_203109:2011, Krivanek_MM20_832:2014, Pohl_U150_16:2015}.
%
%
%
%
%
%
%
%
\section{CONCLUSION}
%
%
In the present work, the fully relativistic description of the twisted 
electron radiative recombination with a bare nucleus was presented.
The interaction of the incident electron with the ionic target was 
taken into account to all orders in $\alpha Z$.
It was done by determining the vortex electron wave function as the 
solution of the Dirac equation in the central field.
The solution was constructed in such a way that its asymptotic has the 
form of the superposition of the free twisted and outgoing spherical 
waves.
The resulting wave function was used for the description of two 
different experimental scenarios, namely with macroscopic and mesoscopic 
targets, beyond the Born approximation.
%
%
\\
\indent
%
%
In the case of the macroscopic target, the comparison of the results, 
which were obtained within the Born approximation and with a usage of the 
developed formalism, has been conducted.
For the sake of comparison clarity, the analytical nonrelativistic 
expressions for both approaches were also considered.
It was found that the total cross section for the $2$ keV vortex 
electron RR into the $1s$ state of the hydrogen atom being calculated 
within the Born approximation differs from the exact value by $23\%$.
This discrepancy increases very rapidly with the growth of the $\nu = 
\alpha Z / p$ parameter and for the recombination with the bare lithium 
nucleus amounts to $53\%$.
Contrary to the cross section, the normalized angular distribution and 
the Stokes parameters being obtained within the Born approximation 
coincide with the exact values in the nonrelativistic case.
In the framework of the relativistic formalism, however, this 
result is no longer valid.
%
%
\\
\indent
%
%
For the macroscopic target it was also found that the linear 
polarization of the emitted photon becomes negative at certain conical 
angles.
This means that the photon is polarized perpendicular to the reaction 
plane.
In the conventional plane-wave case, the degree of linear polarization is strictly 
positive, i.e., the photon is polarized in the reaction plane.
%
%
\\
\indent
%
%
Additionally, the situation when the incident electron is a coherent 
superposition of two vortex states with different $m$ was studied.
In this case, the asymmetry of the angular distribution was calculated.
It was found that the asymmetry becomes most pronounced at $\Delta m = \pm 1$ and 
decreases rapidly with the growth of $\Delta m$.
The analytical nonrelativistic expression for the angular distribution 
was also presented.
%
%
\\
\indent
%
%
For the mesoscopic target the dependence of the total cross section on 
the distance between the target center and the twisted electron 
propagation direction has been investigated.
The dependence of the Stokes parameters on the target position has been 
also studied.
It has been found that both the total cross section and the Stokes 
parameters are sensitive to the spatial structure of the incoming 
electron state, i.e. to $m$.
However, this dependence vanishes with the growth of the target size.
%
%
\\
\indent
%
%
At the end, let us add that the developed formalism can be utilized for 
the description of other processes involving twisted electrons and heavy 
ionic or atomic targets beyond the Born approximation.
%
%
%
%
%
%
%
%
\section*{ACKNOWLEDGEMENTS}
%
%
The authors are grateful to A.~I.~Milstein for useful discussions.
This work was supported by RFBR (Grants No.~16-02-00334, 
No.~16-02-00538, and No.~15-02-05868), and SPbSU (Grants 
No.~11.38.269.2014 and No.~11.38.237.2015).
VAZ acknowledges financial support from the government of St. Petersburg.
%
%
%
%
%

%
\end {document}